\documentstyle[10pt,emulateapj]{article}

\received{April 12 2001}

\begin{document}

\title{The X-ray background from groups and clusters of galaxies}

\author{Xiang-Ping Wu and Yan-Jie Xue}

\affil{National Astronomical Observatories, Chinese Academy
                 of Sciences, Beijing 100012, China}

\begin{abstract}
We present an estimate of the X-ray background (XRB) spectrum 
from the warm-hot intergalactic medium (IGM) associated with groups 
and clusters of galaxies, using purely the observationally determined 
X-ray luminosity function (XLF) and X-ray luminosity 
($L_{\rm x}$) - temperature ($T$) relations for groups and clusters.
As compared with previous semi-analytic models based on 
the Press-Schechter mass function,  our approach  
provides a much simpler and more realistic way to evaluate 
the XRB from groups and clusters in the sense that  we make no 
assumption about the dynamical and heating properties of the IGM, and 
the intrinsic dispersion in the  $L_{\rm x}$-$T$ relations  
due to different physical mechanisms among different groups and clusters
can be also included. 
It shows that the resulting 0.1-10 keV XRB spectrum by summing up 
the X-ray emission from all groups and clusters is consistent with
current upper limits placed on the contribution from diffuse gas
to the XRB. This may have profound implications for our understanding of
the missing baryons in the universe. 
\end{abstract}


\section{Introduction}

The missing baryon problem still has no resolution in sight at present.
It is commonly believed that  most of the baryons in the universe  
exist in the form of diffuse warm-hot intergalactic medium (IGM) 
at temperatures of $T\sim10^5$--$10^7$ K (Cen \& Ostriker 1999),
which may contribute a non-negligible fraction of the soft X-ray
background (XRB) as a result of thermal bremsstrahlung emission.
Therefore, the measurement of the soft XRB constitutes a critical test 
for the presence of the warm-hot IGM and also 
for the models of structure formation.
Important progress in the issue has been made 
over the past few years (e.g. Pen 1999; Phillips, Ostriker \& Cen 2000; 
Wu, Fabian \& Nulsen 2001; Dav\'e et al. 2001; 
Voit, Evrard \& Bryan 2001). Essentially, most of the warm-hot IGM 
associated with the gravitational potentials of less massive dark halos 
must reside outside of the systems characterized by their virial radii. 
Otherwise, the XRB produced by the gravitationally heated and 
bound IGM in groups and poor clusters 
will greatly exceed the upper limits set by current 
X-ray observations. This implies that the preheating of IGM either by
non-gravitational processes such as supernovae,
AGNs and galactic winds or by purely gravitational processes due to 
large-scale density perturbations must play a potentially important role in 
the early phase of structure formation, which raises the IGM entropy
and makes the IGM harder to compress.
Such a scenario is supported by both the discovery of the entropy 
excess in groups relative
to that can be achieved in the pure accretion shock heating (Ponman,
Cannon \& Navarro 1999) and the remarkable departure of the 
observed X-ray luminosity ($L_{\rm x}$) - temperature ($T$) relations 
for groups and clusters
($L_{\rm x}\propto T^{3-5}$) (Xue \& Wu 2000 and references therein)
from the simple gravitational scaling law ($L_{\rm x}\propto T^2$)
(Kaiser 1986).

While a sophisticated estimate of the XRB produced by the warm-hot IGM
must rely upon cosmological  hydrodynamic simulations
(e.g. Phillips et al.  2000), 
semi-analytic models, which emphasize the essential physics behind
the problem, provide a simple approach to understanding quantitatively 
the issue, in a complementary manner to hydrodynamic simulations.
Indeed, the pioneering work of Pen (1999) based on the cosmic virial
theorem and the standard Press-Schechter (PS) mass function 
has already revealed the 
properties of the expected XRB from the gravitationally bound IGM 
within the virialized systems, in gross consistency with 
subsequent numerical results. In particular, the employment of the observed 
X-ray luminosity - temperature relations for groups and clusters 
in the prediction of the XRB spectra 
permits the inclusion of other heating effects on the IGM, 
regardless of the details of heating mechanisms and processes 
(Kitayama, Sasaki \& Suto 1998; Wu et al. 2001).  
In this paper, we conduct an alternative, 
semi-analytic approach to the estimate of the XRB, 
using purely the observationally determined quantities such as 
the X-ray luminosity function (XLF) 
of groups and clusters and their X-ray luminosity  - temperature relations.
The advantages of this method are 
as follows:  First, it provides a much simpler and also more 
realistic way to estimate the contribution of
the warm-hot IGM contained in groups and clusters to XRB, in the sense that 
the influence of non-gravitational heating on the IGM can be naturally 
included; Second, it allows one to correctly remove the component  
from known population of groups and especially clusters in the 
measurement of XRB, which will be of potential importance  
in search for the missing baryons residing in large-scale structures;
Finally, it will be interesting to compare the XRB expected 
from the XLF and $L_{\rm x}$-$T$ relations of groups and clusters with 
other independent observational constraints such as
the current XRB surveys by {\sl ROSAT}, {\sl ASCA}, {\sl Chandra} and
{\sl XMM}. Of course, the accuracy and reliability of this 
semi-analytic method depend on our knowledge of X-ray 
groups and clusters revealed by current X-ray observations.
We have recently applied the same method to the study of
the Sunyaev-Zeldovich cluster counts, and found a noticeable difference 
between the prediction  by the PS mass function and that by
the XLF of clusters (Xue \& Wu 2001). 
Throughout this paper we assume a flat cosmological  
model ($\Lambda$CDM)
of $\Omega_{\rm M}=0.3$, $\Omega_{\Lambda}=0.7$ and $h=0.68$.

\section{$L_{\rm x}$-$T$ relation} 

The X-ray luminosity - temperature relation is a good indicator of 
the dynamical and heating properties
of the IGM associated with the underlying gravitational potentials
of dark halos.  It follows that the bolometric X-ray luminosity
of groups and clusters
should obey the simple gravitational scaling law 
$L_{\rm x}^{\rm bol}\propto T^2$, if there is no other heating
mechanism (Kaiser 1986).  However, it has been well known that 
the observationally determined $L_{\rm x}^{\rm bol}$-$T$ relations
on group and cluster scales deviate significantly from this simple scaling,
with $L_{\rm x}^{\rm bol}\propto T^{3-5}$
(e.g. Edge \& Stewart 1991; David et al. 1993; Wu, Xue \& Fang 1999;
Xue \& Wu 2000 and references therein). This is often interpreted as being
due to the preheating of IGM by supernova explosions and/or AGNs before
the IGM falls into the dark matter potential of groups and clusters 
(e.g. Kaiser 1991; Cavaliere, Menci \& Tozzi 1997; Balogh, Babul 
\& Patton 1999; but see Bryan 2000). Consequently, 
the employment of the observed $L_{\rm x}$-$T$ relations 
in the estimate of the XRB from the IGM in groups and clusters may allow us 
to include the non-gravitational heating effect.

We determine the $L_{\rm x}$-$T$ relations for groups and clusters
using the non-exhausted catalog of X-ray groups and clusters 
compiled by Wu et al. (1999) and Xue \& Wu (2000). The updated 
catalog contains 55 groups and 191 clusters whose X-ray 
temperature and luminosity are both available, and the corresponding
$L_{\rm x}$-$T$ relation in the 0.5-2.0 keV band is shown in the upper
panel of Figure 1.
In order to properly take the intrinsic dispersion of the 
$L_{\rm x}$-$T$ distribution into account in our estimate of XRB below,
we smooth the observed data points  in the following way:
For any given X-ray luminosity $L_{\rm x}$, we construct a subsample of
ten neighbor groups/clusters that have the minimum value of
$|L_{\rm x,i}-L_{\rm x}|$. The average temperature of
the ten groups/clusters weighted 
by their measurement uncertainties ($\Delta T_i$) is used as 
the central temperature $T$ at $L_{\rm x}$: 
$T=\sum (T_i/\Delta T_i^2)/\sum (1/\Delta T_i^2)$. 
The temperature scatter around the central temperature $T$ is
described by the standard deviation of the ten data points 
$(T_i,i=1,\cdot\cdot\cdot,10)$ from $T$. 
The reconstructed $L_{\rm x}$-$T$  relation over a broad X-ray luminosity 
range of $4.5\times10^{40}<L_x<6.8\times10^{44}$ erg s$^{-1}$ 
in the 0.5-2.0 keV band is displayed in the lower panel of Figure 1. 
Nevertheless,  at the lower and higher X-ray luminosity ends
beyond the above range the best fit $L_{\rm x}$-$T$ 
relations for groups and clusters will be used, which read
\begin{eqnarray}
kT=10^{0.41\pm0.13}L_{\rm x}^{0.19\pm0.06}\, & {\rm groups};\\
kT=10^{0.73\pm0.05}L_{\rm x}^{0.41\pm0.01}\, & {\rm clusters},
\end{eqnarray}
where $L_{\rm x}$ and $kT$ are in units of 
$10^{44}$ erg s$^{-1}$ (in the  0.5-2.0 keV band) and keV, respectively. 
The two relations intersect at $L_{\rm x}=3.5\times10^{42}$ erg s$^{-1}$, 
and show no apparent evolution at least out to $z\approx0.8$
(e.g. Mushotzky \& Scharf 1997; Della Ceca et al. 2000).

\section{XLF}

Essentially, we follow an approach similar to our recent 
application of the XLF of clusters for the expectation of the 
Sunyaev-Zeldovich cluster counts (Xue \& Wu 2001). 
We model the differential XLF of groups/clusters 
by the Schechter function (e.g. Ebeling et al. 1997;
Collins et al. 1997; Rosati et al. 1998; 2000; Burke et al. 1997;
De Grandi et al. 1999):
\begin{equation}
\frac{dn}{dL_{\rm x}}=A\exp(L_{\rm x}/L^*_{\rm x})L_{\rm x}^{-\alpha},
\end{equation}
We consider both the non-evolving and evolving XLFs of groups and clusters,
which can be characterized by   
$A=A_0(1+z)^{\bar{A}}$ and $L_{\rm x}^*=L_{\rm x0}^*(1+z)^{\bar{B}}$.
For the non-evolving XLF ($\bar{A}=0$ and $\bar{B}=0$), 
we adopt the local XLF constructed by Ebeling et al. (1997) in 
the 0.5-2 keV band: 
$A_0=3.32_{-0.36}^{+0.33}\times10^{-7}$ Mpc$^{-3}$ 
($10^{44}$ ergs s$^{-1}$)$^{\alpha-1}$,
$L_{\rm x0}^*=5.70^{+1.29}_{-0.93}\times10^{44}$ ergs s$^{-1}$ and 
$\alpha=1.85^{+0.09}_{-0.09}$, where the Hubble constant is $h=0.5$.
For the evolving XLF, we use the evolution parameters given by 
Rosati et al. (2000) for an Einstein-de-Sitter universe: 
$\bar{A}\approx0$ and $\bar{B}=-3$. 
We also adopt a combined model or partially evolving XLF 
suggested recently by Gioia et al. (2001),
in which the non-evolving XLF given by Ebeling et al. (1997) 
is applied to all groups and clusters except for the luminous
clusters beyond $z=0.3$ and with 
$L_{\rm x}\geq1.8\times10^{44}$ erg s$^{-1}$ in the 0.5-2 keV band. 
For the latter we choose $\bar{A}\approx0$ and $\bar{B}=-3$.
We convert the above XLFs and X-ray luminosity $L_{\rm x,0}$
in the Einstein-de-Sitter universe into the ones 
in the $\Lambda$CDM model by demanding that the observed number 
of groups/clusters in a given redshift interval ($z,z+dz$) be conserved 
and by utilizing the relation
\begin{equation}
L_{\rm x}=\left[\frac{D_L}{D_{L,0}}\right]^2 L_{\rm x,0},
\end{equation}
where $D_L$ and $D_{L,0}$ are the corresponding luminosity distances. 
Note that the current XLFs are only valid down to  
$L_{\rm x}\approx1\times10^{42}$ erg s$^{-1}$ in the 0.5-2 keV band.
Nevertheless, in order to test how significant the X-ray emission from 
the warm gas associated 
with elliptical galaxies and small groups affects the soft XRB, 
we will make an attempt to extrapolate the current XLFs to 
$L_{\rm x,min}=1\times10^{40}$ erg s$^{-1}$.

\section{XRB: expectation}

Integrating the X-ray emission of all the groups and clusters 
with X-ray luminosity above the threshold $L_{\rm x,min}$
and over redshift space yields the total XRB intensity at a given 
frequency $\nu$: 
\begin{equation}
J(\nu)=\int\int \left(\frac{dL_{\rm x}/d\nu}{4\pi D_{\rm L}^2(z)}\right)
       \left(\frac{dn(L_{\rm x},z)}{dL_{\rm x}}\right)
       \left(\frac{dV}{d\Omega dz}\right) dL_{\rm x}dz,
\end{equation}
where $dV$ is the comoving volume element: 
\begin{equation}
\frac{dV}{d\Omega}=\frac{c}{H_0}\frac{(1+z)^2}{E(z)}D_{\rm A}^2 dz,
\end{equation}
$D_{\rm L}(z)$ and  $D_{\rm A}(z)$ are 
the luminosity and angular diameter distances to groups/clusters at $z$,
respectively. 
Since the $L_{\rm x}$-$T$ relation and XLFs of groups/clusters are 
both given in the 0.5-2 keV band, we adopt the optically thin and 
isothermal plasma emission model with a metallicity of $Z=0.3Z_{\odot}$ 
by Raymond \& Smith (1977) to convert the X-ray luminosity 
in the 0.5-2 keV band into the X-ray luminosity
per unit frequency, $dL_{\rm x}/d\nu$, in which line emission is also included.

We numerically integrate equation (5) 
for the three different evolutionary scenarios of XLF. It turns out
that the difference in the expected 
XRB between  the non-evolving and evolving XLFs of groups/clusters is only 
minor especially when the scatter of the observationally 
determined $L_{\rm x}$-$T$ relations is included (see Figure 3).
Essentially, the non-evolving XLF only leads to a slightly 
larger XRB flux than the evolving or partially evolving models.
Note that the current evolutionary models only admit  
the pure luminosity evolution of clusters. 
In Figure 2 we display the resulting 0.1-10 keV XRB spectra 
for the partially evolving XLF,  in which   
the total XRB is also decomposed into the contributions of groups and clusters
separated at $L_{\rm x}=3.5\times10^{42}$ erg s$^{-1}$ in the 0.5-2 keV band 
and at different redshift ranges.
It is apparent that the predicted hard XRB above $\sim1$ keV is dominated by 
clusters at intermediate reshifts $0.1<z<1$, while most of the soft XRB 
is produced by groups, as was naturally expected. 
In particular, nearby groups within $z=0.1$ make little 
contribution to the soft XRB, in contrast to the distant groups 
beyond $z=1$ which can give arise to a large fraction of the soft XRB 
below $\sim0.4$ keV.

\section{XRB: observational constraints}

We now compare our predicted XRB flux from the IGM in groups and clusters 
with the upper limits set by current observations.

A considerably large fraction of the hard (2-10 keV) XRB has been resolved 
into discrete sources, which contribute a total sky brightness of  
$(1.3-1.75)\times10^{-11}$ erg s$^{-1}$ cm $^{-2}$ deg$^{-2}$ 
for flux down to $1.4-2\times10^{-15}$ erg s$^{-1}$ cm$^{-2}$ 
(Mushotzky et al. 2000; Giacconi et al. 2001; Hasinger et al. 2001).
The new result from 300 ks exposure of the {\sl Chandra} Deep 
Field South gives a  value of 
$(1.46\pm0.20)\times10^{-11}$ erg s$^{-1}$ cm$^{-2}$ deg$^{-2}$
(Tozzi et al. 2001).
Combined with the 2-10 keV background of 
$1.6-2.3 \times10^{-11}$ erg s$^{-1}$ cm$^{-2}$ deg$^{-2}$ 
(Gendreau et al. 1995; Marshall et al. 1980; Vecchi et al. 1999),
the maximum admitted range of the unresolved flux is 
$(0-1.0) \times10^{-11}$ erg s$^{-1}$ cm$^{-2}$ deg$^{-2}$.

In the soft (1-2 keV) band, the total contribution of the resolved
sources for flux greater than 
$2\times10^{-16}$ erg s$^{-1}$ cm$^{-2}$ amounts to  
$3.38-3.55 \times10^{-12}$ erg s$^{-1}$ cm $^{-2}$ deg$^{-2}$ 
(M$^{\rm c}$Hardy et al. 1998; Hasinger et al. 1998; 
Mushotzky et al. 2000; Giacconi et al. 2001; Tozzi et al. 2001).
The percentage contribution from the discrete and other sources 
to the soft XRB depends  upon the uncertainty on  
the evaluation of the total soft XRB which varies from the lowest value 
$3.7 \times10^{-12}$ erg s$^{-1}$ cm $^{-2}$ deg$^{-2}$ 
(Gendreau et al. 1995), the mid-range value
$4.2 \times10^{-12}$ erg s$^{-1}$ cm $^{-2}$ deg$^{-2}$ 
(Miyaji et al. 1998), to the largest value
$4.4 \times10^{-12}$ erg s$^{-1}$ cm $^{-2}$ deg$^{-2}$ 
(Chen, Fabian \& Gendreau 1997). 
We adopt two additional upper limits to the soft XRB from 
diffuse IGM given by Bryan \& Voit (2001):
$1.8 \times10^{-12}$ erg s$^{-1}$ cm $^{-2}$ deg$^{-2}$ and
$6.8 \times10^{-13}$ erg s$^{-1}$ cm $^{-2}$ deg$^{-2}$ 
for the 0.5-2 keV and  0.1-0.4 keV bands, respectively. 
These values are derived by
combining the XRB intensity measurements (Gendreau et al. 1995;
Barcons, Mateos \& Ceballos 2000) and
the deep surveys with {\sl ROSAT} and {\sl Chandra} (Hasinger et al. 1998;
Giacconi et al. 2001), along with a proper modeling of the
spectral slope. 
Finally, the ``remaining'' fluxes recently discovered by Kuntz, Snowden \& 
Mushotzky (2001) from the {\sl ROSAT} All-Sky Survey 
after peeling off various known X-ray foreground and background components
can also be used as useful constraints on the XRB from the IGM in groups
and clusters, although these remainders may be contaminated by
the Galactic halo emission. 
It should be pointed out that the faint, diffuse X-ray emission from 
groups and small clusters may also be included in the current resolved 
1-2 keV XRB. This arises from the difficulty of clearly separating
X-ray emission between nuclear and extended sources at very faint 
flux level. Moreover, X-ray emission from normal galaxies has become 
detectable in the recent deep exposure by {\sl Chandra} and 
{\sl XMM} (Mushotzky et al. 2000; Giacconi et al. 2001; Hasinger et al. 2001; 
Tozzi et al. 2001).
Therefore, the current residual soft XRB after the resolved 
sources are removed may not strictly be used as an upper limit on 
the X-ray emission from diffuse IGM associated with elliptical galaxies, 
groups and clusters.

We first perform an integration of the total XRB in terms of equation (5) 
over four different energy bands, using again the partially evolving XLF,     
and compare our predictions with the existing limits (Table 1).
Note that in the 0.5-2 keV band,
we can obtain the total XRB flux simply by integrating equation (5)
without the employment of the $L_{\rm x}$-$T$ relations.  
Essentially, our predicted results in all the four bands are well within 
the upper limits placed on the contribution from diffuse gas to the XRB. 
We then compare our expected XRB spectrum from the IGM in groups and clusters
with the observational constraints (Figure 3).
For the latter, we use the mean strength of the upper limits over 
different energy bands except for the data of  Kuntz  et al. (2001).
In order to demonstrate how the scatter in the $L_{\rm x}$-$T$ distribution 
(see Figure 1) affects our  predictions, we employ the Monte Carlo 
technique to determine the uncertainty ($68\%$ confidence limits) in the
resulting XRB. 
It appears that the overall XRB produced by the IGM associated with groups
and clusters is roughly consistent with the current upper limits.
Namely,   the X-ray emission from groups and clusters may account for 
the remaining XRB flux reported by the advanced X-ray detectors 
({\sl ROSAT}, {\sl ASCA}, {\sl Chandra})  after the discrete  
sources are removed.  Note that the upper limits given by Kuntz et al. (2001) 
at $\sim0.2$ keV significantly exceed our predictions, which can be  
attributable to the contribution from the Galactic halo emission.

 \begin{table*}
 \vskip 0.2truein
\caption{XRB: predictions vs observational limits.}
 \vskip 0.2truein
 \begin{tabular}{ccc}
 \tableline
 \tableline
Energy range  & Upper limit$^a$ & Prediction$^a$ \\
 \tableline
2 - 10 keV &  0 - 21    & 1.1 - 1.9 \\
1 - 2  keV &  0.31 - 2.1 & 0.8 - 1.1 \\
0.5 - 2 keV &  3.7         & 2.4 \\
0.1 - 0.4 keV & 1.4        & 1.4 - 5.2 \\
 \tableline
 \end{tabular}
 \parbox {8.5in}{$^{a}$ In units of keV s$^{-1}$ cm$^{-2}$ sr$^{-1}$}
  \end{table*}

We have also compared our predictions with the simulated XRB by Phillips
et al. (2000) (Figure 3). For the latter, the contributions from nearby, 
massive clusters ($z\leq0.2$ and $M\geq5.5\times10^{14}h^{-1}M_{\odot}$) 
were removed. This last restriction will not significantly alter 
our predictions because the effect of the nearby,  massive clusters 
on the total XRB is rather small. 
It appears that there is good agreement between 
our analytical results, based on the XLF and $L_{\rm x}$-$T$ relations of
groups and clusters, and their numerical simulations over the entire
energy band.

\section{Discussion and conclusions}

We have provided an alternative, semi-analytic model 
to the estimate of the contribution from
the warm-hot IGM associated with groups and clusters to 
the XRB, based purely on 
the observationally determined $L_{\rm x}$-$T$ relations and XLF of 
groups and clusters. 
This has enabled us to naturally include both gravitational and 
non-gravitational heating influence
without detailed knowledge of the heating processes.  
The resulting 0.1-10 keV XRB spectrum is roughly consistent
with the upper limits set by current X-ray observations
on the contribution from diffuse gas to the XRB 
(e.g. Hasinger et al. 1998; 2001; Mushotzky et al. 2000; Giacconi et al. 2001;
Tozzi et al. 2001). 
It is thus possible that the residual flux, after the discrete sources are 
removed from the total XRB, is (at least partially) due to the X-ray 
emission from the warm-hot IGM associated with groups and clusters.
If confirmed, this would have profound implications for our understanding
of the missing baryons and their distributions.

As compared with previous semi-analytic models which essentially employ the 
PS mass function and the mass - temperature relation in terms of
virial theorem, combined with either an oversimplified scenario for the IGM  
distribution in dark halos (e.g. Pen 1999) or the  $L_{\rm x}$-$T$ relation
for clusters (Kitayama et al. 1998; Wu et al. 2001),
our semi-analytic approach provides a more straightforward and realistic way to
calculate the XRB from the warm-hot IGM associated with groups 
and clusters. Indeed,  we have made no assumption about the
dynamical and heating properties of the IGM, and included
the intrinsic dispersion, if any, in the  $L_{\rm x}$-$T$ relations 
due to different physical mechanisms among different groups and clusters.
The uncertainty in our predictions, aside from the dispersion of  
the $L_{\rm x}$-$T$ distribution, thus arises mainly from  
the XLF of groups and clusters: Firstly, we have extrapolated 
the current XLF and its evolutionary model to $z>1$ where the XLF 
evolution of clusters is poorly constrained.
Our main prediction would remain unchanged only if the evolutionary scenario 
shown by current observations 
(e.g. Rosati et al. 2000; Gioia et al. 2001; references therein)
is correct, i.e., the bulk of clusters  exhibit no significant evolution 
out to $z\sim1$ and the evolution only commences for the most luminous 
clusters, and if such a scenario should also be applicable to groups.
On the other hand, we have shown that the major uncertainty in 
the predicted XRB arises from the scatter of the $L_{\rm x}$-$T$ relation
rather than from the current available evolutionary models of XLF.  
At this point, it is unlikely that the XBR can be used for the purpose
of testing the evolution of XLF for groups and clusters.
Second, we have extrapolated the current XLF to less massive  
systems with X-ray luminosity down to 
$L_{\rm x}\approx1\times10^{40}$ erg s$^{-1}$ in the 0.5-2 keV band. 
This limit even allows us to
include the contribution from the X-ray emission of elliptical galaxies,
which would affect the estimate of the 0.1-1 keV XRB.
Unfortunately,  there has been no observational justification for 
such an extrapolation, except for the consistency between
our prediction and the numerical result by Phillips et al. (2000).
Third, we have adopted a less vigorous approach to converting the XLF 
in an Einstein-de-Sitter universe to that in $\Lambda$CDM model. 
A sophisticated treatment of the problem is to reconstruct the XLF
of groups and clusters and its evolutionary model 
in the $\Lambda$CDM model from real data, and then proceed to  
the computation of the XRB from the IGM in groups and clusters, as we
have done for the $L_{\rm x}$-$T$ relation.

Overall, the XRB spectrum in the 0.1-10 keV band 
estimated from the available $L_{\rm x}$-$T$ 
relation and XLF of groups and clusters is  consistent with  
the residual XRB reported by current observations after the source
contributions are removed. This would be of great significance for
future detection of 
the missing baryons which are believed to exist in the
form of the warm-hot IGM associated with poor clusters and
groups that are further embedded in large-scale structures.

\acknowledgments
We gratefully acknowledge the constructive suggestions by an anonymous 
referee that greatly improved the presentation of this work. 
This work was supported by
the National Science Foundation of China, under Grant No. 19725311
and the Ministry of Science and Technology of China, under Grant
No. NKBRSF G19990754.


\clearpage

\figcaption{X-ray luminosity ($L_{\rm x}$) - temperature ($T$) 
relation ($\Lambda$CDM model) in the 0.5-2 keV band  
for groups and clusters. Upper panel: Raw data (55 groups and 191 clusters); 
Lower panel: Smoothed $L_{\rm x}$-$T$ relation. 
\label{fig1}}

\figcaption{The XRB spectra predicted by the XLF and 
$L_x$-$T$ relations of groups and clusters. The energy resolution
is taken to be $\Delta\log E=0.02$. Upper panel: Contributions of groups 
($1\times 10^{40}$ erg s$^{-1}$ $\leq L_x\leq 3.5\times 10^{42}$ erg s$^{-1}$)
and clusters ($L_x\geq 3.5\times 10^{42}$ erg s$^{-1}$);
Lower panel: Contributions of groups and clusters at different redshift ranges.
\label{fig2}}

\figcaption{Comparison of the predicted XRB by groups and clusters
(shadowed region) with the observational constraints on the contribution  
of diffuse gas to the XRB.   
Dotted line: the upper limit in the 2-10 keV band 
(Mushotzky et al. 2000; Giacconi et al. 2001; Tozzi et al. 2001). 
Three regions filled with different symbols ($\cdot$, $ + $, $\times$)  
in the 1-2 keV range correspond to three evaluations of 
the total XRB intensity by
Chen et al. (1997), Miyaji et al. (1998) and Gendreau (1997),
respectively, after the source contributions detected by
Hasinger et al. (1998), Mushotzky et al. (2000),  Giacconi et al. (2001)
\& Tozzi et al. (2001) 
are removed. Filled circles and diamonds are the upper limits derived by
Bryan \& Voit (2001) in the 0.5-2 and 0.1-0.4 keV bands, respectively. 
Filled squares are the remaining fluxes  detected by Kuntz et al. (2001)
after the known X-ray foreground and background components are
removed but the Galactic halo emission may be included.
The numerically simulated XRB spectrum from the warm-hot IGM 
by Phillip et al. (2000) is also illustrated by solid line.
\label{fig3}}

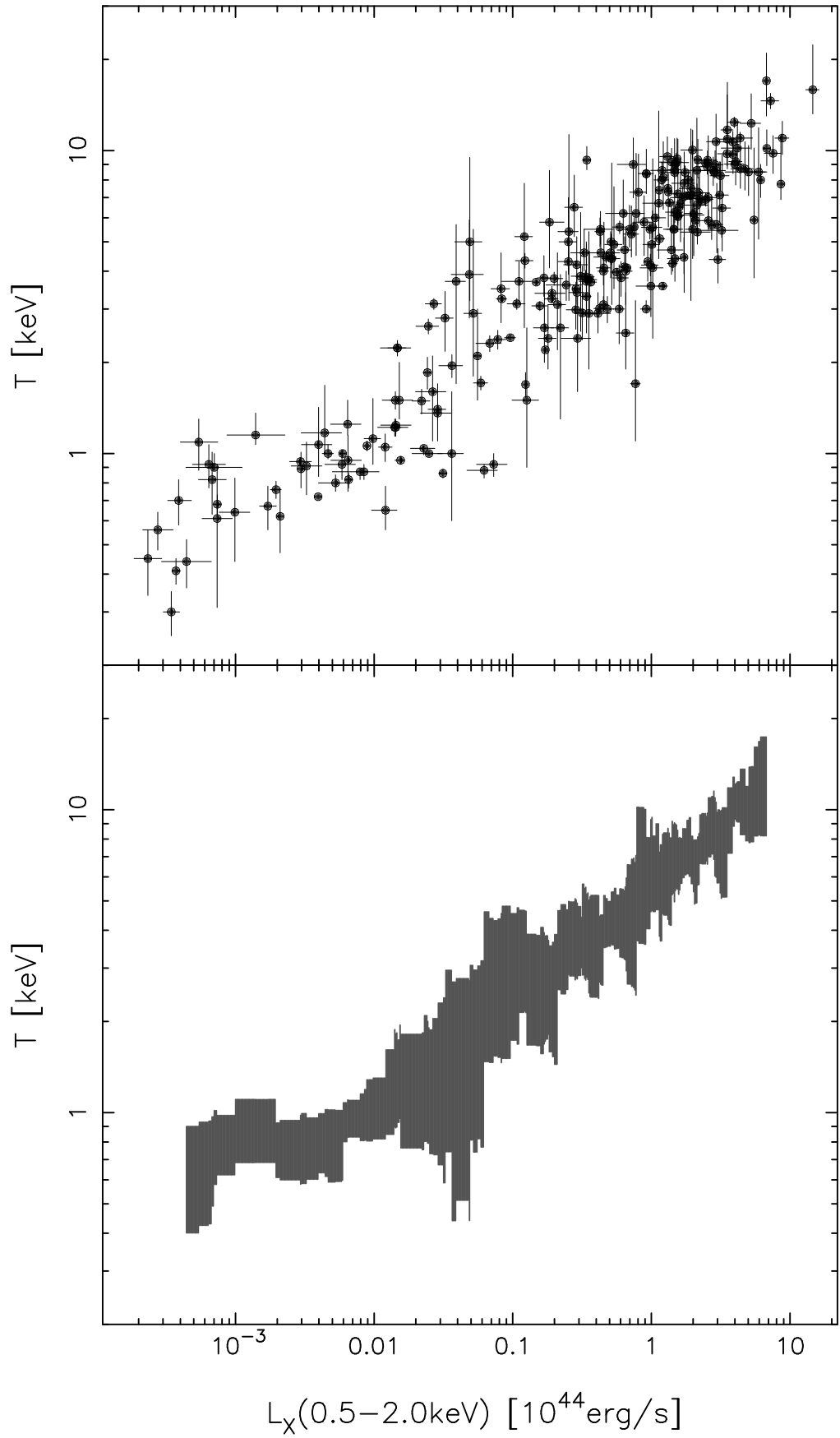
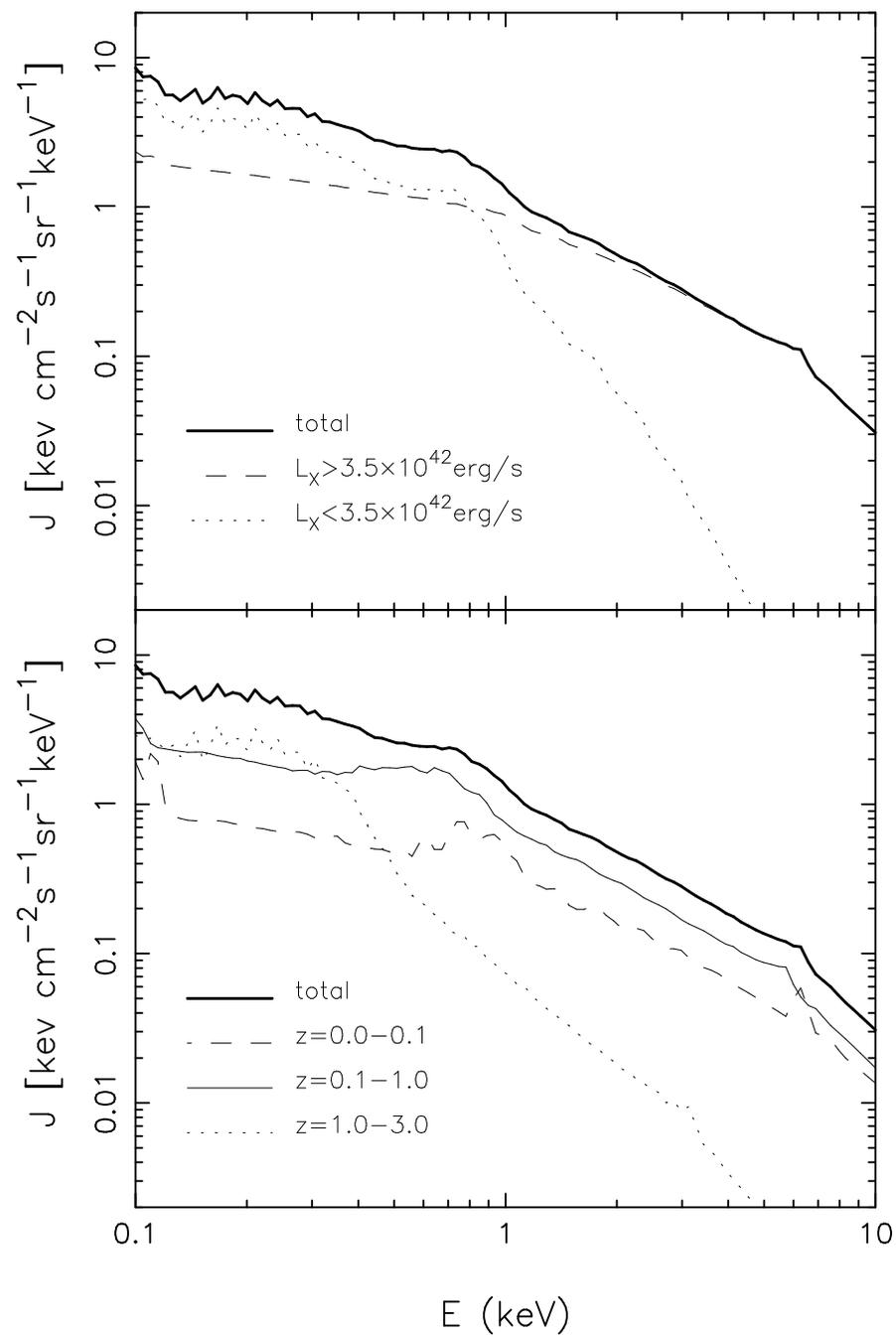
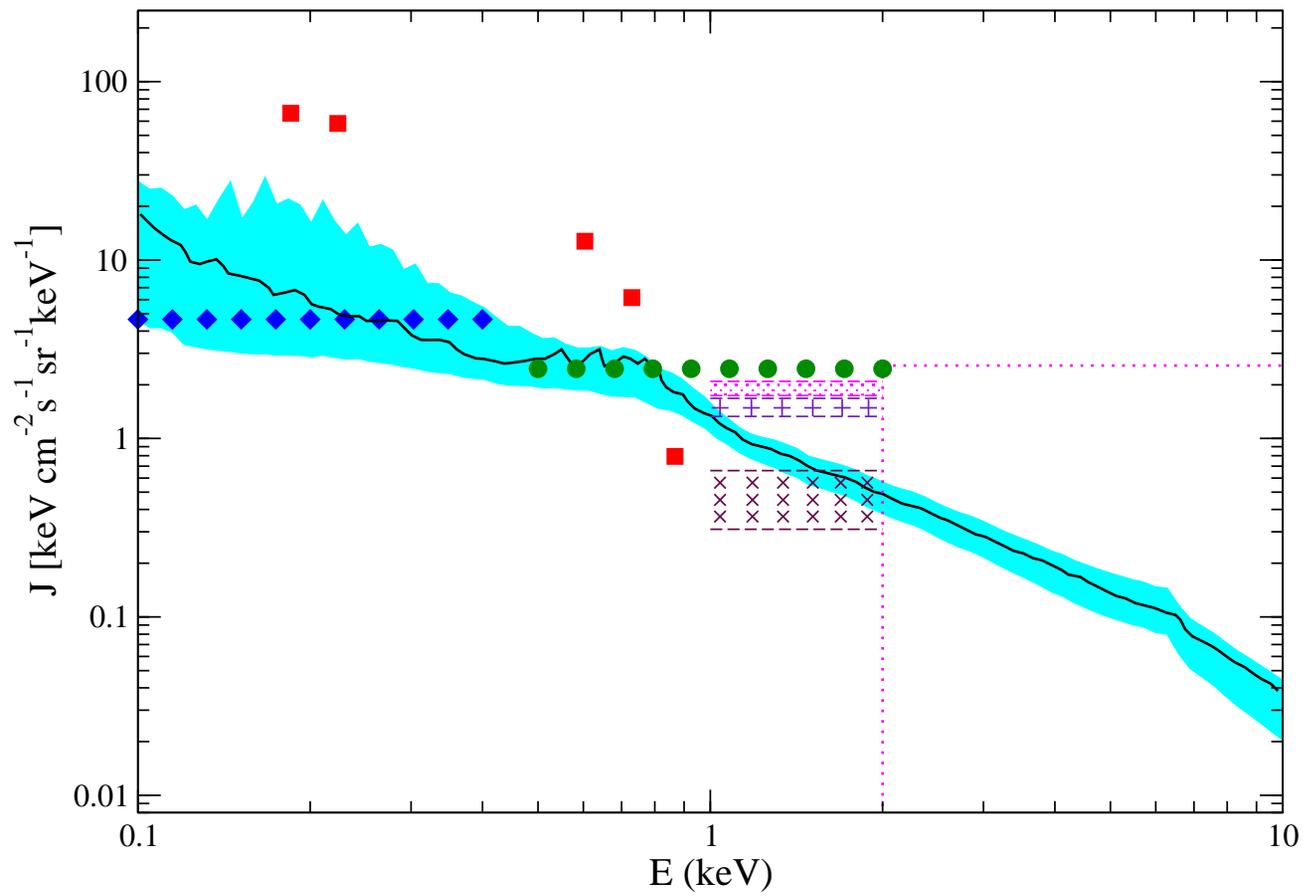
\end{document}